\documentclass{article}
\usepackage{hyperref}

\title{\textbf{On f(R) gravity}}
\author{\textbf{Ahmed Alhamzawi}  \\ \href{ahmzawi@gmail.com}{\emph{ahmzawi@gmail.com}}
 \and \textbf{Rahim Alhamzawi} \\ \href{ralhamzawi@yahoo.com}{\emph{ralhamzawi@yahoo.com}}}

\begin{document}

\maketitle

\begin{abstract}
A review of the new of the problem of dark energy using modified gravity approach is considered. An explanation of the difficulties facing modern cosmology is given and different approaches are presented. We show why some models of gravity may suffer of instabilities and how some are inconsistent with observations.
\end{abstract}

\section{Introduction}
Shortly after the discovery of expansion of the universe different ideas were proposed to try explain the cause of this apparently mysterious expansion. One way to solve this problem is to introduce a new form of energy component, however this approach is plagued with difficulties because this new form of energy has never been detected or observed. Instead of introduced a new form of energy, one can modify the Einstein's theory of GR in such a way to explain accelerated universe. Although this approach has provided many insights to the nature of gravity and has introduced several useful models to explain this problem, at the present there are no fully realized and empirically viable model of modified gravity that explains the expanding universe and satisfies the experimental constrains. The modified gravity models try to tweak Einstein's model of gravity to fit with experimental data, in an attempt to marry up observations with predications. Although the modified gravity model requires no need to invoke a new type of exotic matter its incredibly hard to do and in many cases fails to work that well.
\newline\paragraph*{}
After Einstein introduced his theory of general relativity he modified his field equation by introduced a term that enabled a solution with time-independent, spatially homogeneous and constant positive curvature called "cosmological constant". The cosmological term balances the the attractive force of gravity by making a repulsive gravitational effect to yield a static solution. However, after the discovery of cosmic expansion and several dynamic cosmological models, the cosmological appeared unnecessary.
\newline\paragraph*{}
Dark energy is form energy that tends to accelerate the expansion of space. Since the 1990s, Dark Energy seems to give the most plausible explanation for the experimental observations. Dark Energy was proposed in two forms, one is the cosmological constant, a constant energy density filling space homogeneously \cite{carroll2001cosmological}, and scalar fields such as quintessence, dynamic quantities whose energy density can vary in time and space. Contributions from scalar fields that are constant in space are usually also included in the cosmological constant. The cosmological constant can be formulated to be equivalent to vacuum energy. Scalar fields change extremely slowly with time, consequently it can be difficult to distinguish it from a cosmological constant.

\section{f(R) Models}
Several f(R) models have been presented through the years and many have suffered from either strong instabilities or they fail to explain the past and future conditions of the universe. Here we will consider the most popular of these models:

\subsection{$f(R)=R+{\alpha}R^{n}$}
This model was first introduced by Starobinsky \cite{starobinskii1979spectrum}. It can be shown that $n=2$ is the only viable option \cite{hwang2001f}. Moreover, this model is consistent with the temperature anisotropies observed by the CMB and provides a viable replacement for scalar field inflation since unlike a scalar field leading to inflation, in this case inflation arises as a consequence of the model. This model works on the fact that while the quadratic term provides an accelerated expansion of the universe, the linear term ensures that inflation end gracefully \cite{de2010f}, the inflationary expansion ends when the quadratic term becomes smaller than the linear term.

\subsection{$f(R)=R-\frac{{\mu}^{4}}{R}$}
This model was introduced in 2004 by Carroll et la \cite{carroll2004}. However, shortly after its publication it was shown that this model suffers from series instabilities \cite{de2010f}. For particular values of $\mu$ this model can be used to model the problem of dark energy. The main reason why this model doesn't work is because $f^{\prime\prime}(R)$ is negative which gives a negative value for the mass of the scalar field. The $\frac{1}{R}$ correction is negligible in comparison with R at the high curvatures of the early universe, and becomes important only in the late universe as $R\rightarrow 0$ \cite{faraoni2008f}.

\section{Signatures of modified gravity}
More generally, one can search for signatures of modified gravity by comparing the history of cosmic structure growth to the history of cosmic expansion. Within GR, these two are linked by a consistency relation. Modifying gravity can change the predicted rate of structure growth, and it can make the growth rate dependent on scale or environment. In some circumstances, modifying gravity alters the combinations of potentials responsible for gravitational lensing and the dynamics of non-relativistic tracers (such as galaxies or stars) in different ways, leading to order unity mismatches between the masses of objects inferred from lensing and those inferred from dynamics in unscreened environments.

\section{Constrains and Difficulties facing Modified gravity}
For an f(R) gravity model to be successful, it is must not only serve the purpose for which it was introduced, but it must also pass the tests imposed by Solar System and terrestrial experiments on relativistic gravity, and it must satisfy certain minimal criteria for viability. An f(R) model is criticized as valid if it passes some certain criteria like having the correct Newtonian and post-Newtonian limit, not suffer from instabilities and ghosts, possess the correct cosmological dynamics and give rise to cosmological perturbations compatible with the data from the cosmic microwave \cite{faraoni2008f}. If a model failed to satisfy even a single one of these criteria is taken as a statement that the theory is doomed.

\section{Dark Matter}
The problem of Dark Matter, often called the 'missing mass', has been one of the most series problems facing astronomers and cosmologists today. For yet undermined reason, dark matter is very dim and non-luminous, which makes it very hard to observe or locate. There are many explanations of the problem of dark energy that have been proposed. The different ideas proposed illustrates how scientists work and how they interpret the data. There are two main ways of explaining Dark Matter, one is proposing unseeing non-luminous bodies and the other is by modifying gravity to fit the experimental data. For a long time the first approach, i.e. non-luminous bodies, has been attractive to physicists and astronomers. Physicists prefer it because the dark matter particle if discovered could be just one of the supersymmetric particles which provides an extension to the standard model. Although many experimentalists donated large portions of their time in search of these particles none has been detected. However, the none detection of these particles doesn't mean the falsification of the non-luminous bodies approach because it can be argued that there are other competing theories that can provide a more plausible solution without the need to invoke mysteries particle with partially unknown properties. Astronomers lean to dark matter because it explains many of the problems facing modern astronomy like rotation curves, anomalous lensing and the growth of large scale structures.

\section{Quintessence}
Quintessence is a scalar field that evolves over time. The scalar field hypothesis is the theory that gives most interesting results. To explore the idea of Quintessence we must define some variables
    \begin{eqnarray}\label{Density Parameter}
        {\Omega}_{i}\equiv{\frac{{\rho}_{i}}{\rho}}
    \end{eqnarray}
KG Equation
    \begin{eqnarray}\label{KG Equation}
        \ddot{{\phi}}+3H\dot{\phi}+\frac{\partial{V}}{\partial{\phi}}=0
    \end{eqnarray}
This is the equation of a particle of unit mass with one-dimensional coordinate $\phi$, moving in a potential $V(\phi)$ with a frictional force $-3H\dot{\phi}$. The field will run towards lower values of $V(\phi)$ that are at local minimum. Of course there is no reason why the values of $V(\phi)$ where it is stationary should be small.

The total energy density $\rho={\rho}_{\gamma}+{\rho}_{\phi}$, where ${\rho}_{\gamma}$ is the matter energy density and ${\rho}_{\phi}$ is the scalar energy density, is defined as
    \begin{eqnarray}\label{Energy density}
        \rho=\frac{{\rho}_{0}}{a^{\alpha}}+\frac{1}{2}{\dot{{\phi}}^2}+V(\phi)
    \end{eqnarray}

using these results and substituting $a=a_{0}t^{P}$, $\phi={\phi}_{0}ln(t)+c$ and $V=V_{0}e^{-\lambda\kappa\phi}$ in (\ref{KG Equation}), the KG equation becomes

    \begin{eqnarray}\label{KG Equation 2}
        \frac{{\phi}_{0}(3P-1)}{t^{2}}-\frac{V_{0}\lambda\kappa{e}^{-\lambda\kappa{\phi}_{0}}}{t^{\lambda\kappa{\phi}_{0}}}=0
    \end{eqnarray}
the equation can be solved by equating the powers of $t$ and substituting ${\lambda\kappa{\phi}_{0}}=2$.
In a matter-dominated universe and a radiation-dominated universe, with scale factors $a(t)=a_{0}t^{\frac{2}{3}}$ and $a(t)=a_{0}t^{\frac{1}{2}}$ respectively, the universe expands, but is not accelerating. There exists a power-law solutions for the scale factor $a(t)$, with a logarithmic scalar field and an exponential scalar potential which results in an expanding universe, but it is not necessarily accelerates.In a universe dominated solely by a scalar field, the universe expands and accelerates if $\lambda>\frac{2{\phi}_{0}e^{2}}{V_{0}\kappa}$. In these simple models, the power of the scale factor does not change with time. Whatever dominates to begin with, will always dominate.
\newline\paragraph*{}
Many models of quintessence have a tracker behavior, which partly solves the cosmological constant problem. In these models, the quintessence field has a density which closely tracks but is less than the radiation density until matter-radiation equality, which triggers quintessence to start having characteristics similar to dark energy, eventually dominating the universe \cite{weinberg2008cosmology}. This naturally sets the low scale of the dark energy. When comparing the predicted expansion rate of the universe as given by the tracker solutions with cosmological data, a main feature of tracker solutions is that one needs four parameters to properly describe the behavior of their equation of state.

\end{document}